\documentclass[twocolumn,prb,superscriptaddress,preprintnumbers,amsmath,amssymb,showpacs,floatfix]{revtex4-1}

\usepackage{graphicx}
\usepackage{dcolumn}
\usepackage{bm}
\usepackage{color}

\newcommand{\led}{$L_{2,3}$}
\newcommand{\lmco}{LaMn$_{0.5}$Co$_{0.5}$O$_3$}

\newcommand{\cvo}{CoV$_2$O$_6$}
\newcommand{\acvo}{$\alpha$-CoV$_2$O$_6$}
\newcommand{\gcvo}{$\gamma$-CoV$_2$O$_6$}
\newcommand{\coz}{Co$^{2+}$}
\newcommand{\cod}{Co$^{3+}$}
\newcommand{\cco}{Ca$_3$Co$_2$O$_6$}
\newcommand{\eg}{$e_g$}
\newcommand{\tg}{$t_{2g}$}

\begin{document}

\title{Spectroscopic evidence for giant orbital moment and magnetic anisotropy induced by local distortions in \acvo}

\author{N.~Hollmann}
    \affiliation{Max-Planck-Institut f\"ur Chemische Physik fester Stoffe, N\"othnitzer Str. 40, 01187 Dresden, Germany}

\author{S.~Agrestini}
    \affiliation{Max-Planck-Institut f\"ur Chemische Physik fester Stoffe, N\"othnitzer Str. 40, 01187 Dresden, Germany}
    
\author{Z.~Hu}
    \affiliation{Max-Planck-Institut f\"ur Chemische Physik fester Stoffe, N\"othnitzer Str. 40, 01187 Dresden, Germany}

\author{Z.~He}
    \affiliation{State Key Laboratory of Structural Chemistry, Fujian Institute of Research on the Structure of Matter, Chinese Academy of Sciences, Fuzhou, Fujian 350002, China}

\author{M.~Schmidt}
    \affiliation{Max-Planck-Institut f\"ur Chemische Physik fester Stoffe, N\"othnitzer Str. 40, 01187 Dresden, Germany}

\author{C.-Y.~Kuo}
    \affiliation{Max-Planck-Institut f\"ur Chemische Physik fester Stoffe, N\"othnitzer Str. 40, 01187 Dresden, Germany}

\author{M.~Rotter}
    \affiliation{Max-Planck-Institut f\"ur Chemische Physik fester Stoffe, N\"othnitzer Str. 40, 01187 Dresden, Germany}

\author{A. A. Nugroho}
    \affiliation{Faculty of Mathematics and Natural Sciences, Institut Teknologi Bandung, Jl. Ganesha 10, Bandung 40132, Indonesia}

\author{V.~Sessi}
    \affiliation{European Synchrotron Radiation Facility, Bo\^ite Postale 220, F-38043 Grenoble C\'edex, France}

\author{A.~Tanaka}
    \affiliation{Department of Quantum Matter, ADSM, Hiroshima University, Higashi-Hiroshima 739-8530, Japan}

\author{N.~B.~Brookes}
    \affiliation{European Synchrotron Radiation Facility, Bo\^ite Postale 220, F-38043 Grenoble C\'edex, France}

\author{L.~H.~Tjeng}
    \affiliation{Max-Planck-Institut f\"ur Chemische Physik fester Stoffe, N\"othnitzer Str. 40, 01187 Dresden, Germany}

\date{\today}

\pacs{}

\begin{abstract}

We present a combined experimental and theoretical study on the local magnetism of the Co ions in the spin-chain compound \cvo, which crystallizes in two different allotropic phases, $\alpha$- and \gcvo. Using x-ray magnetic circular dichroism, we have found a very large and a moderate orbital contribution to the magnetism in \acvo\ and \gcvo, respectively. Full-multiplet calculations indicate that the differences in the magnetic behavior of $\alpha$- and \gcvo\ phases originate from different local distortions of the CoO$_6$ octahedra. In particular, the strong compression of the CoO$_6$ octahedra in \acvo\ lead to a strong mixture of \tg\ and \eg\ orbitals which, via the local atomic Coulomb and exchange interactions, results in an exceptionally large orbital moment.

\end{abstract}

\maketitle

Low-dimensional magnetic spin systems have attracted high interest in solid state physics due to their complex magnetic behavior. Magnetic ions in these materials are arranged in, e.g. planes, ladders, or linear chains.
The low dimensional magnetic sublattice may lead to a rich phase diagram, where small changes
in magnetic field and temperature induce a rearrangement of the magnetic order. An example is \cco,
in which chains of \cod\ ions are arranged on a triangular lattice with a step-like magnetization
in the ordered state.\cite{aasland97a, kageyama97a, maignan00a, hardy04a} X-ray magnetic dichroism
experiments and theoretical work provided detailed information on the orbital occupation, the charge and spin state,
as well as the orbital moment explaining the magnetic anisotropy in \cco.\cite{burnus06a,wu05a}

Recently, a new spin-chain compound \cvo\ was synthesized.\cite{he09a} The compound crystallizes
in two different allotropic phases, monoclinic \acvo\ and triclinic \gcvo.\cite{jasper84a, mocala85a, mueller91a} The V ions are in a non-magnetic V$^{5+}$ ($3d^0$)
state and the magnetic properties are determined by the Co$^{2+}$ ions. A step-like behavior is found
in the isothermal magnetization curves in the ordered state, both for $\alpha$- and
 \gcvo.\cite{he09a, kimber08a, kimber11a, lenertz11a, yao12a} The saturation magnetizations
 of the two phases, however, are quite different: 4.5 $\mu_B$ and 2.9 $\mu_B$ per Co ion for
the $\alpha$ and $\gamma$ phase, respectively. Presupposing a spin moment of 3~$\mu_B$ for \coz\ ($S=3/2$),
the magnetism for \acvo\ must have a large orbital contribution. Kim \emph{et al.}\cite{kim12a}
 have used density functional theory (DFT) to estimate the orbital moment of both phases. The difference in the magnetic properties between the two phases is attributed to the different local distortion of the CoO$_6$ octahedra. GGA+spin-orbit coupling (SO) approximation, however, underestimates the orbital moment for the $\alpha$ phase. In order to reproduce the experimental moment, it is necessary to include an orbital polarization (OP) term yielding an orbital moment of 1.8~$\mu_B$ with a large uniaxial anisotropy along the $c$ axis.
 The orbital occupation was given in terms of complex orbitals $d_{\pm 2}$, $d_{\pm 1}$, and $d_0$ in analogy to \cco.
 For this compound, one minority electron occupies the $d_{\pm 2}$ orbitals giving an orbital moment close to 2~$\mu_B$.\cite{burnus06a,wu05a}
In case that $d_{\pm 2}$, $d_{\pm 1}$, and $d_0$ are indeed eigenstates for \acvo, the orbital moment could be even 3~$\mu_B$.

However, the local coordination of Co in \acvo\ is different from that in \cco: the magnetic HS Co$^{3+}$ ion in \cco\
has a trigonal prismatic local symmetry, while \acvo\ contains distorted
octahedra exhibiting a different ligand field level scheme. This requires in our opinion a different quantitative explanation of the local magnetic properties. In fact, there are more reasons why \cco\ should not be considered an analogue of \cvo. The latter contains only divalent Co as magnetic ion, while the former has high spin Co$^{3+}$ as magnetic ion and additional low spin Co$^{3+}$ ($3d^6$) as non-magnetic ion. In \cco, the Co chains consist of alternating magnetic and non-magnetic \cod\ ions with the easy axis pointing along the chain direction
and the magnetic exchange coupling works indirectly over the oxygen ligands.
In \acvo, the easy axis is perpendicular to the chain direction and magnetic
exchange is realized by superexchange between the \coz\ ions in their edge-sharing octahedra.

Here we report on our study of the local magnetism in both phases using x-ray magnetic circular dichroism
(XMCD) spectroscopy at the Co-\led\ edge. The XMCD technique provides separate information on spin and orbital
moments, and in addition the x-ray absorption spectroscopy (XAS) line shape yields details on the local electronic
structure and the ligand field splittings. We use a full-multiplet configuration-interaction
approach to model the experimental spectra as well as the experimentally determined magnetic moments. We are able to track down quantitatively and qualitatively the essential ingredients for the anomalously large orbital moment in \acvo. Our results directly confirm that the orbital
magnetism in \acvo\ stems from the unusual local coordination, and that the large size of the moment needs a many-body description.

\begin{figure}[t]
\includegraphics[angle=270, width=0.45\textwidth]{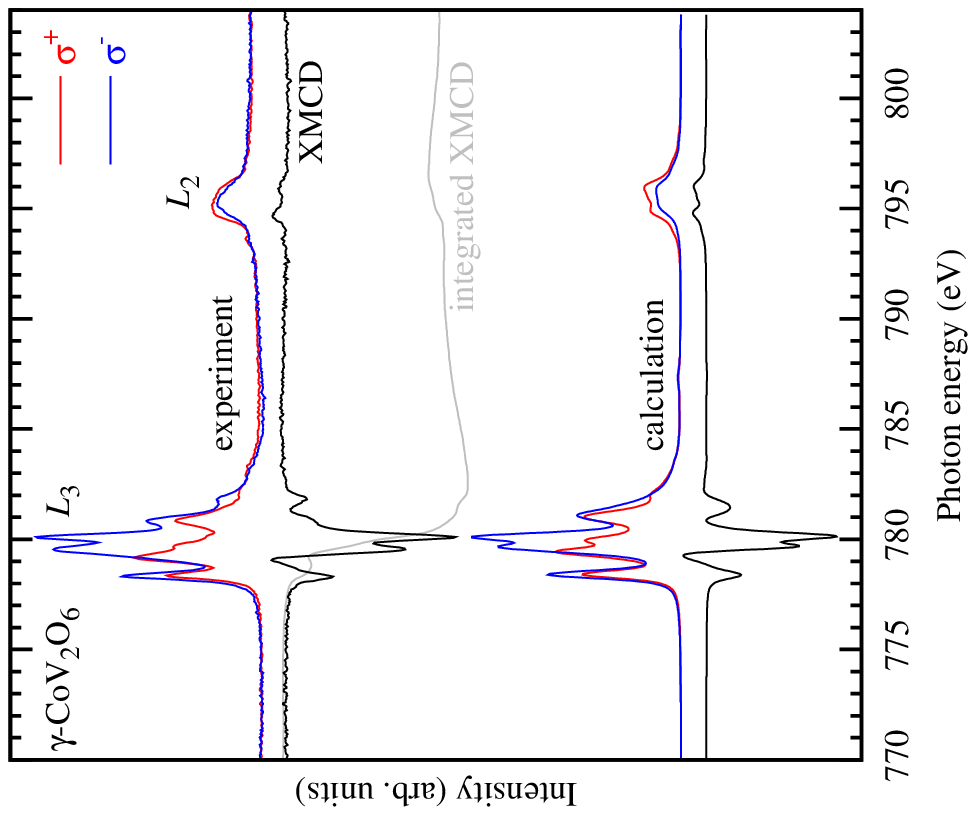}
\caption{\label{fig:gcvo} (Color online) Co-\led\ edge x-ray magnetic circular dichroism for \gcvo.}
\end{figure}

Single crystals of \acvo\ have been grown by the flux method described in Reference~\onlinecite{he09a}. Single crystals of \gcvo\ have been grown by starting from CoC$_2$O$_4\ast$2H$_2$O and V$_2$O$_5$ the stoichiometric mixture was heated in a corundum crucible in air up to $600^{\circ}$C in 24~h and annealed at this temperature for 92~h.
The crystals of \gcvo\ were obtained by chemical vapour transport. The transport process was carried out from a microcrystalline sample of \gcvo\ in an evacuated quarz tube in a temperature gradient from $730^{\circ}$C (source) to $630^{\circ}$C (sink). TeCl$_4$ (2.5~mg/ml) was used as transport agent.
 The XAS experiments were performed at the ID08 beamline of the ESRF synchrotron facility in Grenoble,
 France. The crystals were cleaved \emph{in situ} in ultra-high vacuum in the low 10$^{-10}$~mbar range
to obtain clean sample surfaces. A fast-switchable high-field magnet was used to obtain
the XMCD signal at the Co-\led\ edge. The energy resolution was $\approx 0.25$~eV with a degree
of circular polarization of larger than 99\%. The magnetic field direction was chosen parallel to the $c$ axis for \acvo, which is the
direction where the step-like behavior is observed in the magnetization measurements.\cite{he09a}
We used a field of $B=5$~T at 10~K, so that approximately 90\%\ of the
saturation magnetization is picked up, according to the magnetization curves in Ref.~\onlinecite{singh12a}.
The magnetic field was parallel to the $b$ axis for \gcvo. A CoO sample was measured simultaneously
and used as reference. The absorption spectra were obtained in total electron yield mode and the minor
self-absorption effects were corrected using the procedure given by Nakajima et al.\cite{nakajima99a}
DFT calculations have been performed with the FPLO code\cite{koepernik99a} for $\alpha$- and \gcvo, using the experimental crystal structures.\cite{markkula12a, mueller91a} We have chosen the LDA approximation\cite{perdew92a} and calculated the density of states on a mesh of 24x24x24 $k$ points.

\begin{figure}[t!]
\includegraphics[angle=270, width=0.45\textwidth]{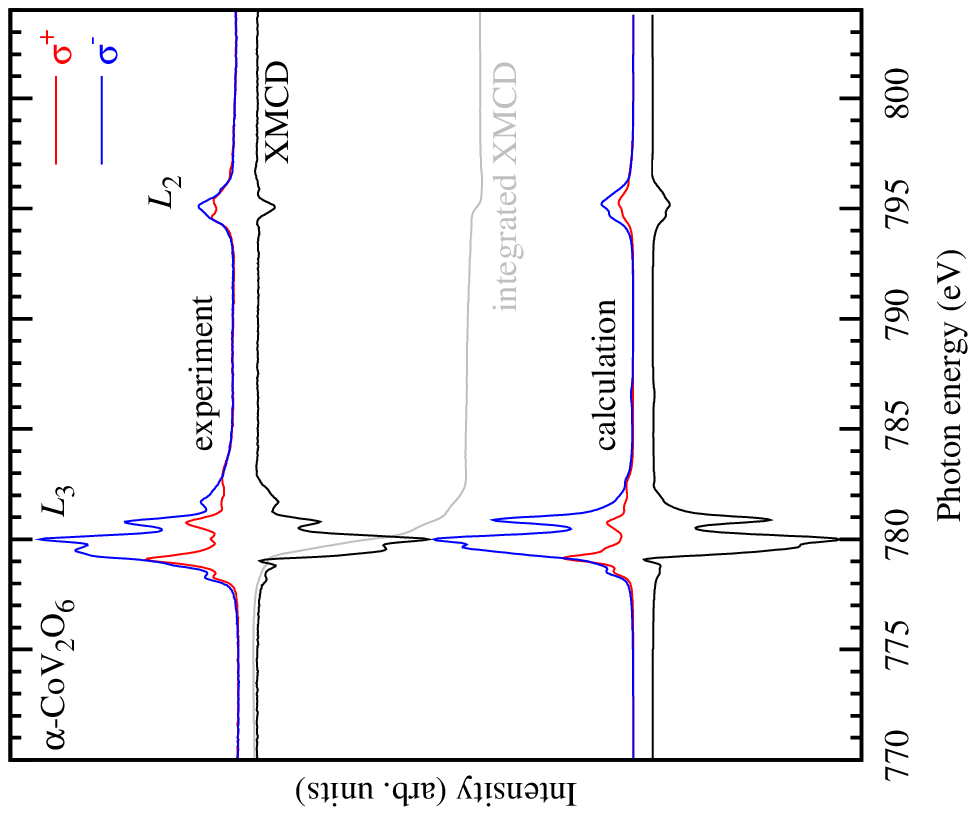}
\caption{\label{fig:acvo} (Color online) Co-\led\ edge x-ray magnetic circular dichroism for \acvo.}
\end{figure}

We first discuss the experimental XAS of \gcvo\ depicted in Fig.~\ref{fig:gcvo}.
The \led\ edge shows the dipole-allowed $2p^63d^n\to 2p^53d^{n+1}$ transition, which is highly sensitive to
the electronic ground state of the $3d$ shell in terms of valence and the local coordination.\cite{degroot94a}
 The lineshape and the energy position of \gcvo\ spectrum clearly indicate that the charge state is \coz.
The XAS lineshape of \gcvo\ is very similar to that of CoO or \lmco\ compounds which have Co ions in an almost regular octahedral environment.\cite{burnus08a}
The XMCD signal, being sensitive to both magnitude and sign of the spin and orbital contributions to the magnetic moment, resembles the one of \lmco\cite{burnus08a}\ in lineshape and opposite sign at the $L_3$
 and $L_2$ lines, indicating similar local magnetic properties,
including a moderate value of the orbital moment of about 1~$\mu_B$. For a quantitative analysis,
we first utilize the sum rules in XMCD.\cite{thole92a} For the ratio of orbital and spin moment, when the magnetic quadrupole moment $T_z$ is ignored,\cite{tzremark} the sum rules can be expressed as\cite{carra93a}
\begin{equation}\label{eq:ratio}
\frac{m_{orb}}{m_{spin}}=\frac{2}{3}\cdot \frac{\int_{L_{2,3}}(\sigma^+-\sigma^-)dE}{\int_{L_{3}}(\sigma^+-\sigma^-)dE-2\int_{L_{2}}(\sigma^+-\sigma^-)dE}.
\end{equation}
The advantage of this sum rule is that the desired values for spin and orbital moment can be extracted directly from the XMCD spectrum without the need to subtract a step-edge background. For \gcvo, it yields $m_{orb}/m_{spin}\approx 0.49$.

The XAS of \acvo\ in Fig.~\ref{fig:acvo} also shows a high spin \coz\ state. The lineshape of $\sigma^+$,
 $\sigma^-$, and the XMCD spectra, however, are very different from those of \gcvo. The most striking feature is
the same negative sign of the XMCD signal at the $L_3$ and $L_2$ lines. This is quite unusual for a solid, as the sign of
the XMCD signal is typically opposite at the $L_3$ and $L_2$ lines reflecting a reduction of orbital moment of
the transition metal ions in the solid state from its atomic value.
The negative signs of the XMCD signal in \acvo\ are in that sense indicating that the Co ions in \acvo\ may effectively have free-ion like quantum numbers in its ground state.\cite{vanderlaan91a} Using the sum rule (\ref{eq:ratio}), $m_{orb}/m_{spin}\approx 0.73$ is found, which is clearly larger than the value for \gcvo.

\begin{figure}[t]
\includegraphics[width=0.45\textwidth]{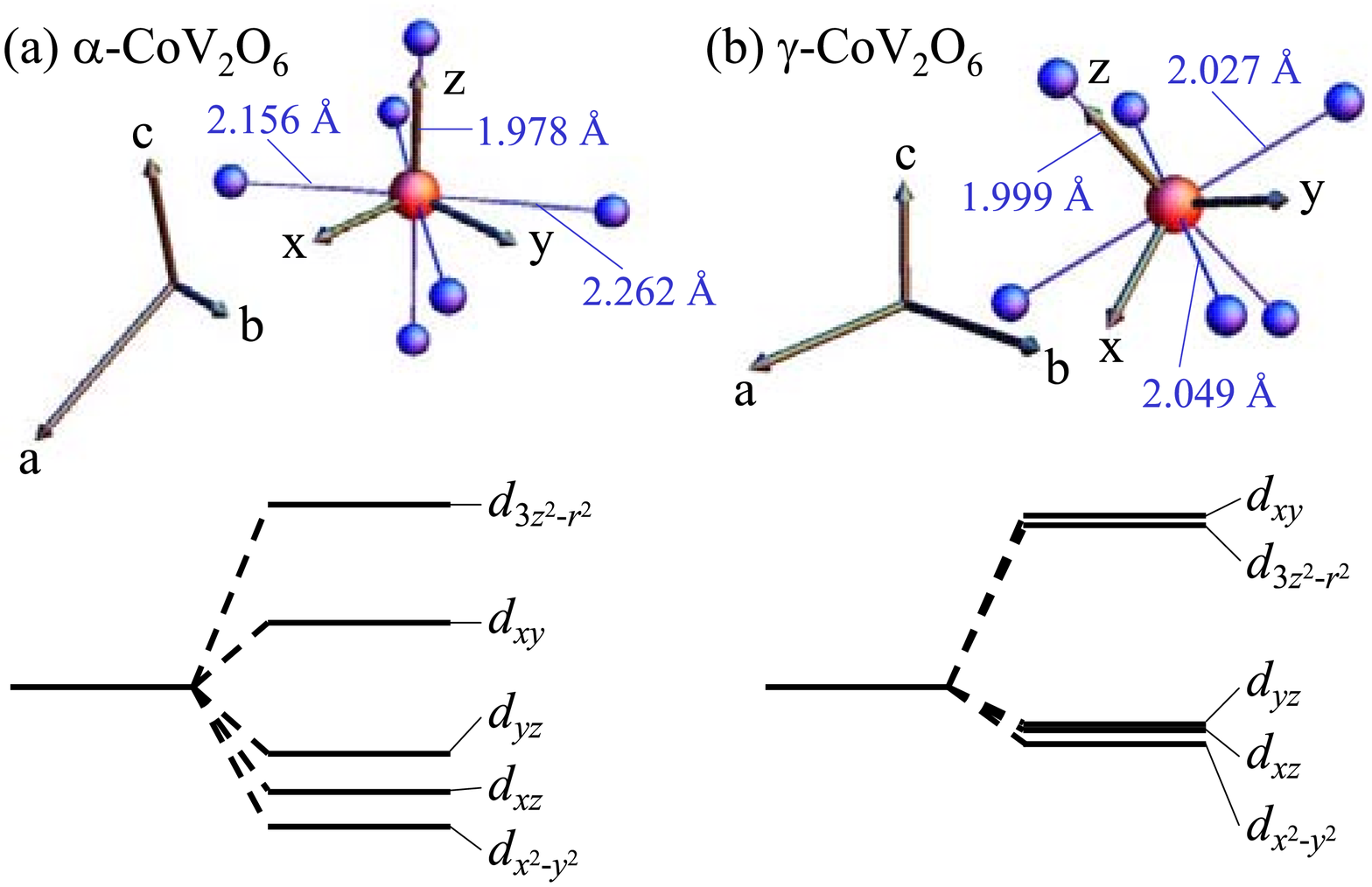}
\caption{\label{fig:ecf} (Color online) Local coordination of the Co ions with Co-O bond lengths. The schematic ligand field splittings are given below, omitting the small mixing between $d_{x^2-y^2}$ and $d_{3z^2-r^2}$ orbitals. }
\end{figure}

As the Co-L$_{2,3}$ XAS lineshape contains signatures of the microscopic origin of the giant orbital moment in \acvo, and the large difference compared to \gcvo, we make a further analysis by carrying out simulations of the spectra using the well-proven full-multiplet configuration-interaction approach.\cite{degroot94a, thole97a} The calculations were performed using the XTLS 9.0 code.\cite{tanaka94a} The monopole parts of the Coulomb interaction as well as the charge-transfer energy\cite{parameters} were taken from CoO.\cite{csiszar05a} Covalence between
Co and O was taken into account using the formalism of Slater and Koster.\cite{slater54a}
Hybridization strengths were calculated from Harrison's book\cite{harrison} using the experimental Co-O bond lengths.\cite{jasper84a, mueller91a} See also the local coordinate system as illustrated in Fig.~\ref{fig:ecf}.
The parameters to be optimized are then the \textit{ionic} one-electron crystal field splittings between the relevant $3d$ orbitals.

As a first step, we tune the values for the ionic part of the crystal fields such that the resulting effective crystal fields or effective one-electron energy levels of the $3d$ orbitals (i.e. including the effect of the hybridization with the oxygens), are close to the estimates from \textit{ab-initio} band structure calculations. We carried out DFT calculations for \cvo\ in the non-magnetic state using the FPLO code and obtained Co $3d$ partial density of states (PDOS) very similar to those of Kim \emph{et al.}.\cite{kim12a}. The first moments of the Co-$3d$-dominated PDOS yield effective crystal field values of $10Dq$= 0.52 eV (splitting between the $e_g$ and $t_{2g}$ orbitals) , $\Delta t_{2g}$=0.06 eV (the splitting of the $d_{x^2-y^2}$ orbital with respect to the $d_{xz}$/$d_{yz}$ orbitals), and $\Delta e_g$=-0.24 eV (lowering the $d_{xy}$ orbital with respect to the $d_{3z^2-r^2}$ orbital) for \acvo. It turned out that the simulated XAS and XMCD spectra do not provide an optimal match with the experimental ones. Also the orbital moment calculated using these parameters is about 1.1~$\mu_B$, i.e. smaller than value deduced using the XMCD orbital sum rule on the \acvo\ experimental spectra.

In the next step we fine tune the crystal field values to obtain an optimal fit to the \acvo\ experimental spectra. The effective crystal field values came out to be $10Dq$= 0.52 eV, $\Delta t_{2g}$=-0.21 eV, and $\Delta e_g$=-0.60 eV and a sketch of the effective one-electron energy levels is displayed in Fig.~\ref{fig:ecf}. The resulting simulations are shown in Fig.~\ref{fig:gcvo}. One can clearly see that these simulations reproduce all the experimental $\sigma^+$, $\sigma^-$, and XMCD spectra very well. The orbital moment calculated using these parameter set is $m_{orb}$=1.9 $\mu_B$. The calculated spin moment is $m_{spin}$=2.5 $\mu_B$, giving a total magnetic moment of 4.4 $\mu_B$ for the $\alpha$ phase, i.e. reproducing excellently the saturation magnetization value,\cite{he09a, lenertz11a} and in good agreement to the result from the sum rule given above. We note that these optimized effective crystal field values do not deviate very much in absolute terms from the above listed band structure estimates. Nevertheless, discrepancies of 0.27 eV in $\Delta t_{2g}$ and 0.36 eV in $\Delta e_g$ are very large compared to the Co $3d$ spin-orbit constant, implying that more accurate methods\cite{streltsov05a, chang09a, wu11a} need to be explored if one wishes to make estimates for the crystal fields on the basis of \textit{ab-initio} theories.

Another important note that we would like to make is that we have calculated not only the XAS and XMCD spectra using the full atomic multiplet theory (i.e. including all spin and orbital flip terms of the atomic Coulomb and exchange interactions), but also the orbital and spin moments. Instead, when we use a one-electron scheme,\cite{smith92a} we will find an orbital moment of only $m_{orb}$=1.0 $\mu_B$, i.e. substantially smaller than the $m_{orb}$=1.9 $\mu_B$ value in the full multiplet approach (using the same one-electron energy levels). This may explain why Kim \emph{et al.}.\cite{kim12a} needed to include the orbital polarization term in their calculations to improve their DFT results.

The \gcvo\ phase contains two different cobalt sites. In average, we find $10Dq$= 1.07 eV, $\Delta t_{2g}$=-0.02 eV, and $\Delta e_g$=0.04 eV as values for the effective crystal fields in order to reproduce the experimental XAS and XMCD spectra, see Figs.~\ref{fig:gcvo}. The moments calculated using these parameters are $m_{orb}$=0.7$\mu_B$ and $m_{spin}$=1.8 $\mu_B$, giving a total moment of 2.5 $\mu_B$, close to the results of the magnetization experiments.\cite{he09a, lenertz11a}. It is clear that these effective crystal field values corresponds to an almost cubic local Co coordination, see the illustration in the right panel of Fig.~\ref{fig:ecf}.

\begin{figure}[t!]
\includegraphics[angle=270, width=0.42\textwidth]{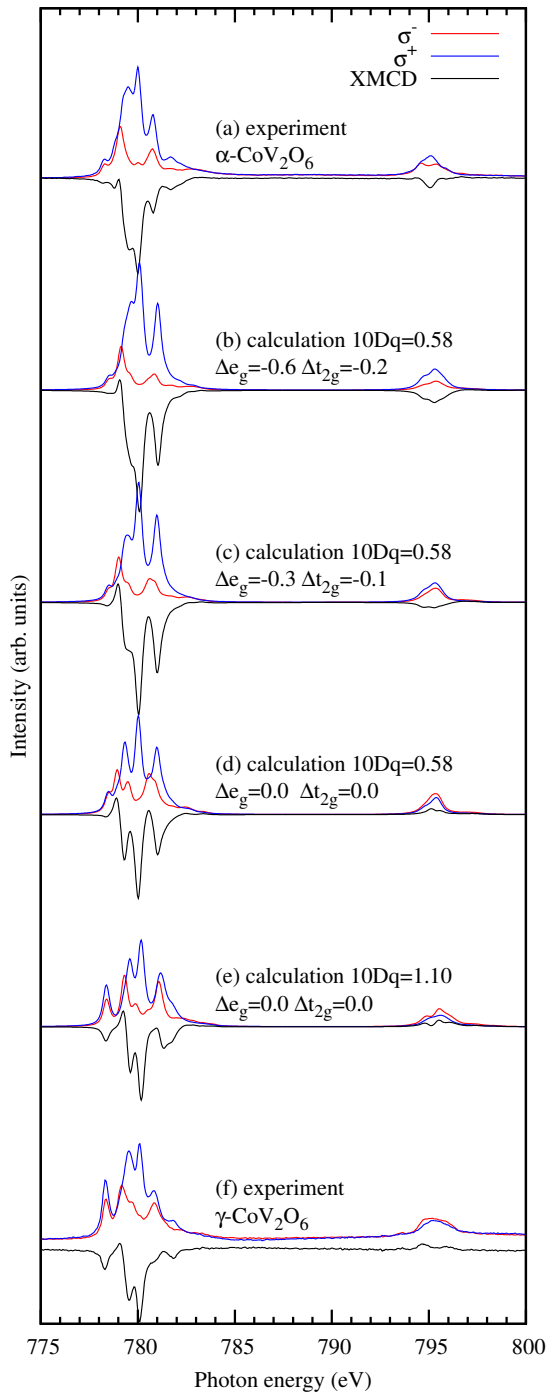}
\caption{\label{fig:sf} (Color online) Influence of octahedral compression on the Co-\led\ edge x-ray magnetic circular dichroism. See text for the details on the calculations.}
\end{figure}

Let us look into the details of the local environment of Co$^{2+}$ to illustrate the reason for the large orbital moment in \acvo\ in a many-body picture. First we regard cubic symmetry and
ignore mixing of \tg\ and \eg\ states. For high spin $3d^7$ with
$t_{2g}^5e_g^2$ occupation, the total spin of $S=3/2$ couples with
the pseudo orbital momentum $\tilde{L}=1$ to $\tilde{J}=1/2$ as
state lowest in energy. The orbital occupation for the hole within
the \tg\ states is somewhat similar to the case of
$t_{2g}^5$,\cite{ballhausen} with a combination of $d_{\pm 1}$ and
$d_{xy}$ orbitals. Due to spin-orbit coupling, the eigenstates are not purely $d_{+1}$ or $d_{-1}$;
the resulting value for the orbital moment is $L_z=2/3$~$\mu_B$. In presence of a tetragonal distortion, depending on whether the octahedron is
compressed or elongated, either the $d_{xy}$ ($b_{2g}$) or the
$d_{\pm 1}$ ($e_g$) manifold is stabilized. If only the \tg\ states are considered, the
value of $L_z$ reaches up to 1~$\mu_B$; this is the maximum that
can be created by one hole occupation of the $d_{+1}$ orbital in \emph{compressed} distortion.
Any in-plane orthorhombic distortion will reduce the orbital moment as it tends to stabilize the hole in one of the real orbitals $d_{xy}$, $d_{xz}$, or $d_{yz}$. To explain the large $L_z$ in \acvo, the \eg\ orbitals have to be taken into account.\cite{kim12a}
The mixing of $d_{xy}$ and $d_{x^2-y^2}$ orbitals brings $d_{\pm 2}$ orbital character into the ground state
and enhances the orbital moment. For a given spin-orbit coupling strength, the amount of the admixture
depends on the type of the CoO$_6$ distortion, namely the energy splitting of the five $3d$ orbitals. There is an additional important difference in the energetics between the $d_{\pm 1}$ and the $d_{\pm 2}$ orbitals: the mixing matrix element due to spin-orbit coupling is twice as large for the $d_{\pm 2}$ orbitals, an effect which is even squared for small energy separations.

From this conjecture, we can conclude that for an enhancement of $L_z$ from a cubic symmetry, the local coordination should have (1) a large compressed distortion to generate strong mixing terms between the $e_g$ and $t_{2g}$ orbitals, (2) a small 10$Dq$ corresponding to large average bond lengths in order to bring the $e_g$ close in energy to the $t_{2g}$ orbitals, and (3) only small to moderate orthorhombic distortion in the $xy$ plane of the octahedron to avoid the formation of real space orbitals.

Let us study a simplified model of the local environment to illustrate its effect on the spectra and the orbital moment. The experimental spectra are given in Fig.~\ref{fig:sf} (a) and (f), for \acvo\ and \gcvo, respectively. To obtain the calculated curves (b)-(e), we have performed ionic calculations and approximated the local environment as tetragonal: the crystal field is characterized by the cubic splitting $10Dq$ between $e_g$ and $t_{2g}$ levels; the levels are further split up by $\Delta t_{2g}$ (lowering the $d_{x^2-y^2}$ orbital with respect to the $d_{xz}$/$d_{yz}$ orbitals) and $\Delta e_g$ (lowering the $d_{xy}$ orbital with respect to the $d_{3z^2-r^2}$ orbital). First compare the spectra in Fig.~\ref{fig:sf} (b)-(d), where the cubic splitting $10Dq$ is held constant while the size of the splittings $\Delta t_{2g}$ and $\Delta e_g$ is reduced from (b) to (d). This resembles a thought experiment in which the average bond length of the CoO$_6$ octahedron is kept constant, while reducing the compression of the apical oxygen atoms. It can be seen in the spectra that the XMCD on the $L_2$ line reverses its sign between curves (c) and (d), which is evidence for the strong dependence of the orbital moment on the amount of compression; the increase of compression when going from (d) to (b) drastically increases the orbital moment. Comparing curve (b) to the experimental spectrum for \acvo\ given in (a), it can be seen that the main features are already properly described in this simple model, and that the line shape of the XMCD reflects the local environment. In curves (d) and (e), the dependence of the spectrum on the cubic splitting $10Dq$ is demonstrated: the line shape of the spectral features for each polarization is changed, but the total integral of the XMCD, corresponding to the amount of orbital moment, is only slightly affected. Increasing the size of $10Dq$ leads to a better resemblance to the experimental spectrum of \gcvo, compare curves (e) and (f). The spectra in Fig.~\ref{fig:sf} demonstrate that indeed the leading term in the difference in the orbital moment between $\alpha-$ and \gcvo\ is the amount of compression of the CoO$_6$ octahedra. For the spectral line shape differences, also the magnitude of $10Dq$ matters.

To summarize, we have examined the local magnetism of the Co spin chain compounds \acvo\ and \gcvo\
experimentally with x-ray magnetic circular dichroism. We are able to confirm quantitatively the large orbital contibution to the magnetic moment for \acvo\
that is caused by the strong compression of the octahedra. That the effect of the local symmetry is exceptionally large can be traced back to the multiplet nature of the atomic Coulomb and exchange interactions within the Co $3d$ shell.

We gratefully acknowledge the ESRF staff for providing beamtime. We thank B.~Kim for fruitful discussions.


\begin{thebibliography}{37}
\bibitem [{tzr()}]{tzremark}%
  \BibitemOpen
  \href@noop {} {}\bibinfo {note} {$T_z$ is usually negligible in octahedral
  environment. In the simulations for the spectra,
  $T_z<0.1$~$\mu_B$.}\BibitemShut {Stop}%
\bibitem [{par()}]{parameters}%
  \BibitemOpen
  \href@noop {} {}\bibinfo {note} {$U_{dd}=6.5$~eV, $U_{pd}=8.2$~eV, and
  $\Delta=6.5$~eV. Slater integrals were reduced by 70\%\ from Hartree-Fock
  values.}\BibitemShut {Stop}%
\end{thebibliography}
\end{document}